\documentclass[preprint,aps,12pt,preprintnumbers,eqsecnum,nofootinbib]{revtex4}
\usepackage{graphicx}
\usepackage{subfigure}

\usepackage{color}
\usepackage{amssymb,amsmath,amsfonts}
\usepackage{epstopdf} 
\usepackage{braket}

\newcommand{\beq}{\begin{equation}}
\newcommand{\enq}{\end{equation}}

\renewcommand{\ol}{\overline}

\newcommand{\x}{\times}
\newcommand{\bd}[1]{{\bf #1}}

\newcommand{\da}{\dagger}
\newcommand{\Tr}{\text{Tr}}

\begin{document}
%
%
\title{\vspace*{0.5in} 
Universal Landau Pole and Physics below the 100 TeV Scale
\vskip 0.1in}
\author{Christopher D. Carone}\email[]{cdcaro@wm.edu}
\author{Shikha Chaurasia}\email[]{scchaurasia@email.wm.edu}
\author{John C. Donahue}\email[]{jcdonahue@email.wm.edu}

\affiliation{High Energy Theory Group, Department of Physics,
College of William and Mary, Williamsburg, VA 23187-8795}
%
%
\date{May 26, 2017}
\begin{abstract}
We reconsider the possibility that all standard model gauge couplings blow up at a common scale in the ultraviolet.  
The simplest implementation of this idea assumes supersymmetry and the addition of a single vector-like generation 
of matter fields around the TeV scale.  We provide an up-to-date numerical study of this scenario and show that either 
the scale of the additional matter or the scale of supersymmetry breaking falls below potentially relevant LHC bounds.  We 
then consider minimal extensions of the extra matter sector that raise its scale above the reach of the LHC, to determine whether 
there are cases that might be probed at a $100$~TeV collider.  We also consider the possibility that the heavy matter sector involves new
gauge groups constrained by the same ultraviolet boundary condition, which in some cases can provide an explanation for the 
multiplicity of heavy states.  We comment on the relevance of this framework to theories with dark and visible sectors.
\end{abstract}
\maketitle

\section{Introduction} \label{sec:intro}
The idea that the three gauge couplings of the standard model may assume a common value at a high energy scale has motivated
a vast literature on grand unified theories~\cite{Langacker:1980js}.  The particle content of the minimal supersymmetric standard 
model (MSSM) is consistent with such a unification, with a perturbative unified gauge coupling obtained around 
$2 \times 10^{16}$~GeV.  However, it was pointed out long ago~\cite{Maiani:1977cg,Cabibbo:1982hy}  that a different framework also 
leads to the correct predictions for the gauge couplings at observable energies, namely one in which the gauge couplings blow up 
at a common scale $\Lambda$ in the ultraviolet (UV):
\begin{equation}
\alpha_1^{-1}(\Lambda)=\alpha_2^{-1}(\Lambda)=\alpha_3^{-1}(\Lambda)=0 \,\,\, .
\label{eq:uvbc}
\end{equation}
Since the SU(3) coupling is asymptotically free, this boundary condition can only be obtained via the introduction of 
extra matter~\cite{Cabibbo:1982hy,Maiani:1986cp,Moroi:1993zj,Andrianov:2013irn}.  Supersymmetric models offer the simplest 
possibility,  a single vector-like generation of mass $m_V$~\cite{Cabibbo:1982hy,Maiani:1986cp,Moroi:1993zj}.  For a chosen value of $m_V$, 
one may fix the scale $\Lambda$ by the requirement that the low-energy value of the fine structure constant $\alpha_{EM}$ is reproduced;  the 
values of  $\sin^2\theta_W$ and $\alpha_3^{-1}$ are then predicted at any chosen renormalization scale $\mu$, up to theoretical uncertainties.  
If a value of $m_V$ can be found in which both $\sin^2\theta_W(m_Z)$ and $\alpha_3^{-1}(m_Z)$ are consistent with the data, then a viable 
solution is obtained.   This approach, followed in Ref.~\cite{Moroi:1993zj}, found $m_V$ around the TeV scale, assuming that $m_V$ is also the 
scale of supersymmetry breaking (which we call $m_{susy}$ below).

A numerical renormalization group analysis cannot directly encode the boundary condition in Eq.~(\ref{eq:uvbc}) since the
gauge couplings are in the non-perturbative regime, where the renormalization group equations (RGEs) cannot be trusted.  In 
Ref.~\cite{Moroi:1993zj}, the boundary condition studied was $\alpha_1(\Lambda)=\alpha_2(\Lambda)=\alpha_3(\Lambda)=10$, 
values that are barely perturbative.  Since the couplings are rapidly increasing as the renormalization scale is increased, one makes 
the reasonable assumption that the value of $\Lambda$  that satisfies this boundary condition is very close to the one given by 
Eq.~(\ref{eq:uvbc}). On the other hand, as the renormalization scale is decreased, the couplings become increasingly perturbative.  
Of particular importance is that the results are insensitive to the precise choice of boundary condition as long as each of the couplings 
is large~\cite{Parisi:1974cf}.   It was shown in Ref.~\cite{Moroi:1993zj}, that varying the $\alpha_i(\Lambda)$ by an order of 
magnitude in either direction has only a small effect on the final results.  We will see this explicitly in our study of the 
one-vector-like-generation scenario in Sec.~\ref{sec:onevgen}.  The insensitivity of the predicted values of $\sin^2\theta_W (m_Z)$ and 
$\alpha_3^{-1}(m_Z)$ to the choice of boundary conditions is due to the existence of an infrared fixed point in the renormalization 
group equation for the ratios of the gauge couplings~\cite{Ghilencea:1997yr}.   Note that this insensitivity includes the case where the 
$\alpha_i(\Lambda)$ are taken to be large but not strictly identical at a common high scale.

The possibility that the gauge couplings may have large values in the UV is interesting from a variety of perspectives.  Large 
couplings may arise in strongly coupled heterotic string theories, which often also provide the additional vector-like states necessary 
to drive the gauge couplings to large values~\cite{Ghilencea:1997yr}.  On the other hand, a universal Landau pole, as defined by 
Eq.~(\ref{eq:uvbc}), may arise in models with composite gauge bosons:  compositeness implies the vanishing of the gauge fields' 
wave-function renormalization factors at the compositeness scale, where the gauge fields become 
non-dynamical~\cite{Eguchi:1974cg}.   Redefining fields and couplings so that the  gauge fields' kinetic terms are always kept in 
canonical form, one finds that the vanishing wave-function renormalization factors translate into the blow-up of the gauge couplings 
at the same scale.  Thus, the framework we study may be consistent with a wider range of possible ultraviolet completions than a 
conventional grand unified theory (GUT) with a large unified gauge coupling, though it is not necessary to commit ourselves to any one of 
them in order to study the consequences at low energies.

An additional motivation relevant to the present work is that the assumption of a universal Landau pole leads to the 
expectation of new physics at a calculable energy scale, $m_V$, that is above the weak scale but potentially within the 
reach of future collider experiments\footnote{This, of course, assumes that the vector-like matter occurs at a single 
common scale.  This assumption is relaxed in Ref.~\cite{Andrianov:2013irn}.}.  In Sec.~\ref{sec:onevgen}, we 
show that the minimal scenario, involving one vector-like generation of additional matter, requires values of either $m_V$ or 
$m_{susy}$ that are below some of the current LHC bounds on vector-like quarks or colored superparticles, respectively.   Although
experimental bounds come with model-specific assumptions that are usually easy to evade, we pursue an 
alternative possibility.  We show that there are small extensions of the new matter sector that successfully reproduce
the correct values of the gauge couplings at $m_Z$ while predicting values of $m_V$ that are above the reach of
the LHC, but below $100$~TeV for some choices of $m_{susy}$.   In some cases, $m_V$ may be light enough
for the vector-like states to be explored at a $100$~TeV hadron collider, which makes study of this sector more interesting.  Aside 
from the presence of the heavy matter fields, one possibility that we also discuss in the present work is that these fields may transform under 
an additional gauge group factor.  The motivation is two-fold:  {\em (1)} By placing the additional matter fields into irreducible representations 
of a new gauge group, we might provide an explanation for the multiplicity of states needed to achieve the desired UV boundary condition.  In the 
case where the heavy matter remains vector-like, the new gauge group can be broken at a much lower scale.   The resulting low-energy theory is 
that of a ``dark" sector consisting of the new gauge and symmetry breaking fields; the heavy matter provides for communication between the dark 
and visible sectors, via a ``portal" of higher-dimension  operators that are induced when the heavy fields are integrated out. The gauge coupling of 
the dark gauge boson is predicted from a boundary condition analogous to Eq.~(\ref{eq:uvbc}) and the magnitude of the portal couplings are 
set by the value of $m_V$ obtained in the RGE analysis. This presents a simpler framework for constraining some of the otherwise free parameters of a 
dark sector than, for example, attempting to embed both dark and visible sectors in a conventional GUT.  {\em (2)} The heavy matter may be chiral under the 
new gauge group.  The structure of the new sector is then more analogous to the the electroweak sector of the MSSM, and the scale $m_V$ 
is associated with  one or more massive gauge bosons that may have observable consequences.

Our paper is organized as follows:  In Sec.~\ref{sec:onevgen}, we consider the consequences of a universal Landau pole in the minimal case 
where the MSSM is augmented by a single vector-like generation.  The study presented in this section differs from the past literature not only in our
use of up-to-date experimental errors for our input parameters, but also in that we allow the scales $m_V$ and $m_{susy}$ to vary independently. In addition,
we consider an alternative choice for the vector-like matter that contributes the same amount to the beta functions at one loop, but differs from
the one-generation scenario at two loops. In Sec.~\ref{sec:next}, we consider extensions of these minimal scenarios, in particular, including a small 
number of additional complete SU(5) multiplets of vector-like matter.  We focus on finding solutions in which $m_V$ is less than $100$~TeV, 
with  a special interest in cases where the vector-like matter is light enough to be detected at a future hadron collider.  In Sec.~\ref{sec:models} we consider 
model building issues associated with the physics at the scale $m_V$, focusing on the implication of additional gauge groups. In Sec.~\ref{sec:conc}, we 
summarize our conclusions.

\section{One vector-like generation} \label{sec:onevgen}

In this section, we consider a minimal scenario studied in the past literature~\cite{Cabibbo:1982hy,Maiani:1986cp,Moroi:1993zj}, the MSSM augmented by an additional 
vector-like generation of matter fields.  We denote the scale of the vector-like matter $m_V$ and we impose the same boundary conditions as in Ref.~\cite{Moroi:1993zj}, namely 
$\alpha_1(\Lambda)=\alpha_2(\Lambda)=\alpha_3(\Lambda)=10$ as an approximation to Eq.~(\ref{eq:uvbc}).  Taking $m_V$ as an input, we determine 
$\Lambda$ by the condition that the weak scale value of the fine structure constant $\alpha_{EM}(m_Z)$ is reproduced.  With $\Lambda$ fixed, we are 
now able to determine the gauge couplings at any lower scale, as a function of our choice for $m_V$.  Above the scale $m_{susy}$, we use the two-loop 
supersymmetric RGEs for the gauge couplings.  Below $m_{susy}$, we do the same using the two-loop nonsupersymmetric RGEs, aside from running 
between the top quark mass and $m_Z$ which we treat as a threshold correction and include at one loop.  We assume the presence of the second
Higgs doublet required by supersymmetry above the scale $m_{susy}$.  Expanding on the approach of Ref.~\cite{Moroi:1993zj}, we do not assume that the scales 
$m_V$ and $m_{susy}$ are the same, though the relaxation of that requirement will only be important in Sec.~\ref{sec:next}.

As indicated in the introduction, the ratios of the gauge couplings are driven towards infrared fixed point values, so that predictions for 
$\sin^2\theta_W$ and $\alpha_3^{-1}$ at $m_Z$ are relatively insensitive to the choice of boundary conditions at the scale $\Lambda$.   For 
example, allowing the $\alpha_i(\Lambda)$ to vary independently between 1 and 100, we find that the their weak-scale values 
scatter within roughly $2\%$ for $\alpha_1(m_Z)$ and $\alpha_2(m_Z)$ and $5\%$ for $\alpha_3(m_Z)$.  Given the same variation of boundary 
conditions, we take the resulting scatter in the values of $\sin^2\theta_W(m_Z)$ and $\alpha_3^{-1}(m_Z)$ as a measure of the theoretical 
uncertainty in our output predictions.   We include these estimates with our numerical results.

The RGEs that we use above the top mass have the form
\beq
\label{eq:rge}
\frac{dg_i}{dt} = \frac{g_i}{16\pi^2}\left[ b_i g_i^2 + \frac{1}{16\pi^2}\left(\sum_{j=1}^3 b_{ij}g_i^2 g_j^2- \!\sum_{j=U,D,E} a_{ij} g_i^2\, \Tr[Y_j Y_j^\da]\right)\right],
\enq
where $t= \ln \mu$ is the log of the renormalization scale, $\alpha_i = g_i^2/ 4\pi$, and the $Y_i$ are Yukawa matrices. The beta function 
coefficients $b_i$ and $b_{ij}$ can be determined using general formulae~\cite{Yamada:1993ga,Machacek:1983tz}.  For example, in the case 
of one vector-like generation with $m_V = m_{susy}$, one finds for $\mu > m_V$
\begin{equation}
b_i = \left(\begin{array}{c} \frac{53}{5} \\  5 \\ 1 \end{array}\right) \,\,\,\,\, \mbox{ and } \,\,\,\,\,
b_{ij} = \left(\begin{array}{ccc} \frac{977}{75} & \frac{39}{5} & \frac{88}{3} \\
\frac{13}{5} & 53 & 40 \\
\frac{11}{3} & 15 & \frac{178}{3} \end{array} \right) , 
\end{equation}   
while for $m_t < \mu < m_V$ we have the nonsupersymmetric beta functions
\begin{equation}
b_i^{NS} =\left(\begin{array}{c} \frac{41}{10} \\ -\frac{19}{6} \\ -7 \end{array}\right) 
\,\,\,\,\, \mbox{ and } \,\,\,\,\,
b_{ij}^{NS}= \left(\begin{array}{ccc} \frac{199}{50} & \frac{27}{10} & \frac{44}{5} \\
\frac{9}{10} & \frac{35}{6} & 12 \\
\frac{11}{10} & \frac{9}{2} & -26 \end{array} \right) .
\end{equation}
More general forms for the one- and two-loop beta functions that take into account the possibility of additional matter are presented 
in Sec.~\ref{sec:next}.  Note that the gauge couplings for $\mu > m_{susy}$ are defined in the dimensional reduction ($\ol{\text{DR}}$) 
scheme, which preserves supersymmetry; the couplings are converted to the modified minimal subtraction scheme ($\ol{\text{MS}}$) 
at the matching scale $\mu=m_{susy}$ before they are run to lower energies.   The gauge couplings in the two schemes are 
related by~\cite{Antoniadis:1982qw}
\begin{equation}
\frac{4\pi}{\alpha_i^{\ol{\text{MS}}}} = \frac{4\pi}{\alpha_i^{\ol{\text{DR}}}} + \frac{1}{3}(C_A)_i \,\,\, ,
\end{equation}
where $C_A= \{0,2,3\} \text{ for } i=1,2,3$.

The coefficients for the terms that depend on the Yukawa couplings in Eq.~(\ref{eq:rge}) are given by
\begin{equation}
a_{ij} = \begin{pmatrix} \tfrac{26}{5} & \tfrac{14}{5} & \tfrac{18}{5} \\ 6&6&2 \\ 4&4&0 \end{pmatrix}  
\,\,\,\,\, \mbox{ and } \,\,\,\,\, a_{ij}^{NS} = \begin{pmatrix} \tfrac{17}{10} & \tfrac{1}{2} & \tfrac{3}{2} \\ \tfrac{3}{2} & \tfrac{3}{2} & \tfrac{1}{2} \\ 2&2&0 \end{pmatrix} \,\,\, ,
\end{equation}
for $\mu > m_{susy}$ and $\mu < m_{susy}$, respectively.  In practice, we only need to take the top quark Yukawa coupling $y_t$ into account, since it is 
significantly larger than the other Yukawa couplings.   Since $y_t$ affects the running of the gauge couplings only through a two-loop term, we
need only include its running at one-loop.  For $\mu > m_{susy}$ we have~\cite{rge}
\begin{equation}
\frac{dy_t}{dt} = \frac{y_t}{16\pi^2}\left( -\sum c_i g_i^2 +6 y_t^2 \right), \quad c_i = \left(\frac{13}{15}, \, 3, \, \frac{16}{3}\right),
\end{equation}
while for $\mu < m_{susy}$~\cite{rge},
\begin{equation}
\frac{dy_t}{dt} = \frac{y_t}{16\pi^2}\left( -\sum c_i ^{\text{SM}}g_i^2 + \frac{9}{2} y_t^2 \right), \quad c_i^{\text{SM}} = \left(\frac{17}{20}, \, 
\frac{9}{4}, \, 8\right).
\end{equation}
For definiteness, we assume $\tan\beta=2$, and compute the weak scale value of $y_t$ via $y_t(m_Z) = \frac{\sqrt 2\, m_t}{v\sin\beta}$, using the 
$\overline{{\rm MS}}$ value of the top quark mass, $160 ^{+5}_{-4}$~GeV~\cite{pdg}, and $v = 246$ GeV.   The value $y_t(\Lambda)$ is computed 
numerically so that we obtain the desired $y_t(m_Z)$ value for a given set of input parameters.   While this approach is sufficient to determine the 
representative impact of including the top quark Yukawa coupling in our RGE analysis, it turns out to be overkill:  in models where the gauge 
couplings blow up in the UV, the top quark Yukawa coupling is rapidly driven to zero in the same limit.  Hence, its effect on the values of $m_V$ 
and $\Lambda$ determined in our numerical analysis turns out to be small, less than the estimates of theoretical uncertainty that we build into 
the analysis.  Although we include it, ignoring $y_t$ altogether does not affect our results qualitatively and can be a useful approach for speeding
up numerical cross-checks.

\begin{figure}[h!]
	\centering
    	\subfigure{\includegraphics[width=0.4\textwidth]{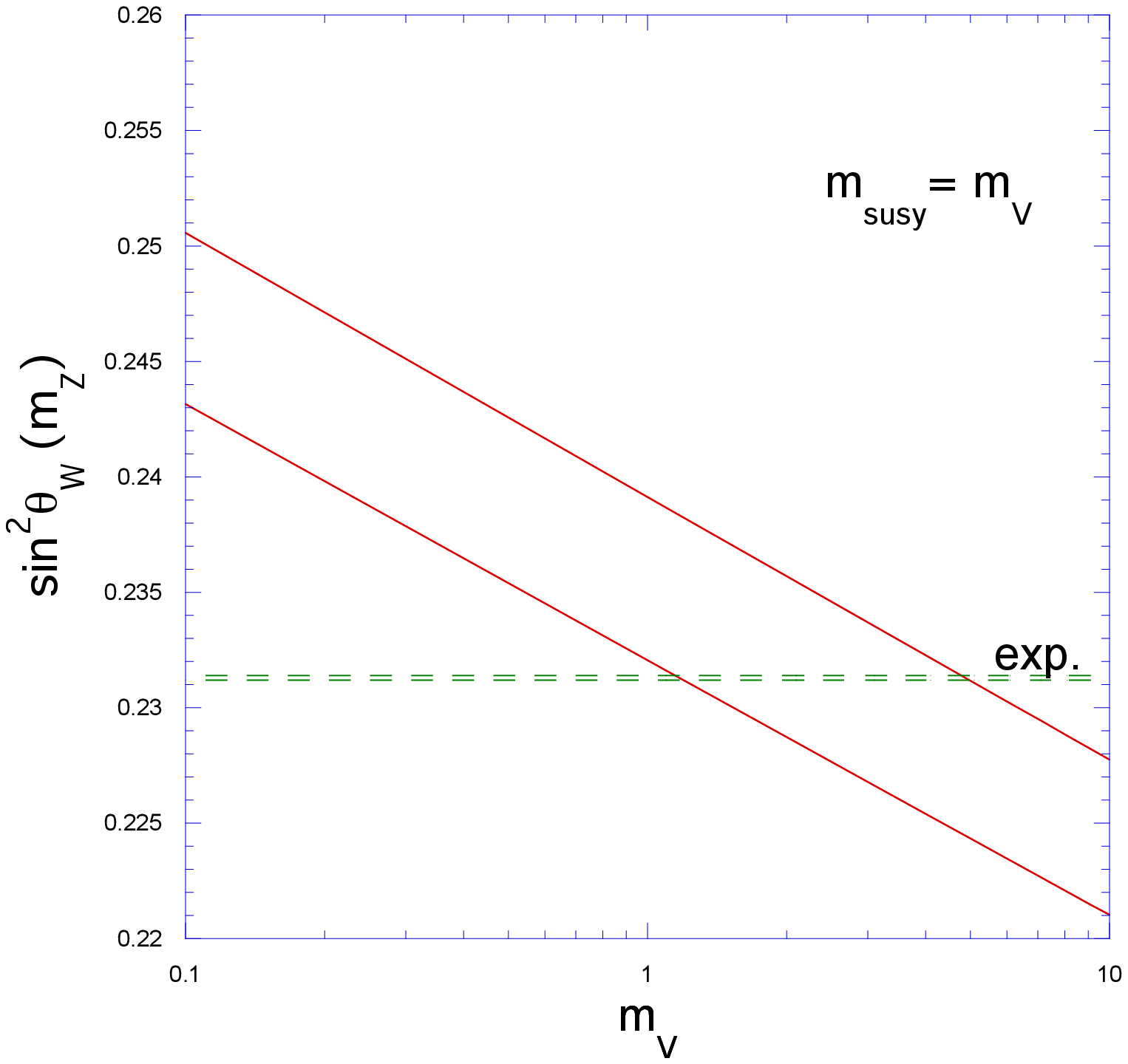}}	\hspace{1em}
	\subfigure{\includegraphics[width=0.4\textwidth]{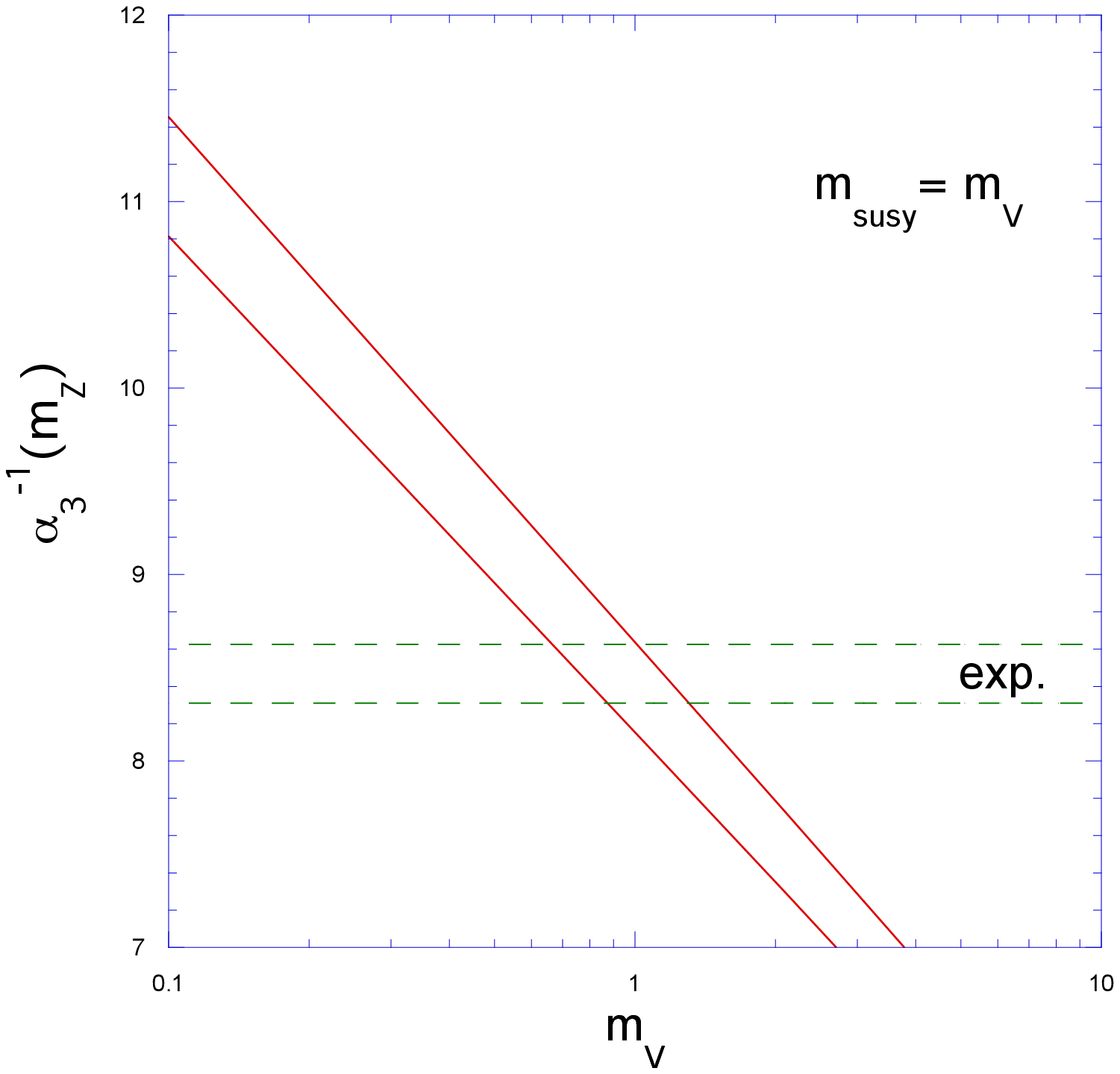}}
	\caption{The dependence of $\sin^2\theta_W(m_Z)$ and $\alpha_3^{-1}(m_Z)$ on the mass of the vector-like generation, $m_V$, including theoretical uncertainties.  In this 
	example, the supersymmetry-breaking scale $m_{susy}$ is identified with $m_V$.  The acceptable ranges of $m_V$ in each of the plots have 
	non-vanishing overlap for $1.15$~TeV$<m_V<1.31$~TeV, indicating a viable solution.}
	\label{fig:5200}
\end{figure}

For a given choice of $m_V$ and $m_{susy}$, the blow-up scale $\Lambda$ is chosen to yield the correct value of the fine structure constant at the weak scale, 
\begin{equation}
\alpha_{\text{EM}}^{-1}(m_Z) = \frac{5}{3} \alpha_1^{-1}(m_Z)+\alpha_2^{-1}(m_Z) ,
\end{equation}
where the factor of $5/3$ comes from the fact that we assume SU(5) normalization~\cite{g1renorm} of the U(1) gauge coupling, as in Ref.~\cite{Moroi:1993zj}.   While 
this makes the analysis compatible with a conventional SU(5) GUT at large coupling, this normalization can also arise directly in string 
theory without an SU(5) GUT~\cite{Dienes:1996du}.  Other normalizations of the U(1) factor are certainly possible, depending on the UV completion.  However, we do
not consider other possibilities here and adopt the normalization that has been assumed almost uniformly in the past literature.  For our numerical study, we take the target central value of 
$\alpha_{\text{EM}}^{-1}(m_Z) = 127.95$~\cite{pdg}.  With $\Lambda$ determined in this way, we compute $\alpha_3(m_Z)^{-1}$ and the Weinberg angle $\sin^2\theta_W(m_Z)$, which is determined by $\alpha_1(m_Z)$ and $\alpha_2(m_Z)$:
 \beq
 \sin^2\theta_W(m_Z) = \frac{3 \alpha_1(m_Z)}{3 \alpha_1(m_Z)+ 5 \alpha_2(m_Z)} \,\,\, .
 \enq
 We compare the output predictions of $\alpha_3(m_Z)^{-1}$ and  $\sin^2\theta_W(m_Z)$, including the theoretical uncertainty that we discussed earlier, to the
 experimentally measured values~\cite{pdg}
 \beq
 \sin^2\theta_W = 0.23129 \pm 5\x 10^{-5}, \quad  \alpha_3^{-1}(m_Z) = 8.4674 \pm 0.0789 \,\,\, ,
 \enq
both given in the $\overline{{\rm MS}}$ scheme.   A previous study of the one vector-like generation scenario found viable solutions with $m_V = m_{susy} \approx 1$ TeV~\cite{Moroi:1993zj}.  Since the time of that work, the experimental errors in $\sin^2\theta_W (m_Z)$ and $\alpha_3^{-1}(m_Z)$ have decreased substantially. Nevertheless, as indicated in Table~\ref{tab:5200}, we find $m_V = m_{susy} \approx 1.2$~TeV, assuming $\pm 2$  standard deviation experimental error bands and using our protocol for determining theoretical error bands; those bands are  both displayed in Fig.~\ref{fig:5200}.   
To determine the theoretical error band, we find the maximum and minimum values of $\sin^2\theta_W(m_Z)$ and $\alpha_3^{-1}(m_Z)$ that are obtained by 
varying the $\alpha_i$ independently between 1 and 100 at the blow-up scale.  In particular, we find that $\sin^2\theta_W(m_Z)$ is maximum when $
\{\alpha_1(\Lambda), \, \alpha_2(\Lambda), \, \alpha_3(\Lambda)\} = \{100, \, 1, \, 100\}$ and minimum when the boundary condition set is $\{1, \, 100, \, 1\}$; $
\alpha_3^{-1}(m_Z)$ is maximized and minimized for the sets $\{100, \, 100, \, 1\}$ and $\{1, \, 1, \, 100\}$, respectively.  We quote the variation in the output 
predictions as a percentage relative to the value obtained when the $\alpha_i(\Lambda) = 10$, for $i=1 \ldots 3$,  in Table~\ref{tab:5200}.  For the values of 
$m_V$ that yield viable predictions for $\sin^2\theta_W (m_Z)$ and $\alpha_3^{-1}(m_Z)$, we find that the scale $\Lambda$ is 
around $8 \x 10^{16}$ GeV.
  
\begin{table}[h!]
\centering
\begin{tabular}{|c|c|c|c|c|c|}
\hline\hline
Model & $m_{\text{SUSY}}$ (TeV) & $m_V$ range (TeV) &  $\Lambda$ range (GeV) & $\alpha_3^{-1}(m_Z)\; \%$ error & $\sin^2\theta_W(m_Z)\; \%$ error
\\ \hline
$(5,2,0,0)$ & $m_V$ & $1.15-1.31 $ & $7.8-8.7 \x 10^{16}$ & $+3.7\%, \; -2.1\%$ & $+1.5\%, \; -1.5\%$
\\ \hline
$(3,2,4,0)$ & $m_V$ & $0.66 -1.16$ & $6.9 - 11 \x 10^{16}$& $+2.8\%, \; -1.5\%$ & $+1.4\%, \; -1.2\%$
\\ \hline\hline
\end{tabular}
\caption{Numerical results for $m_V$ and $\Lambda$ in the one-generation scenario, the (5,2,0,0) model, and a model whose
vector-like sector consists of four $\bd 5+\ol{\bd 5}$ pairs, the (3,2,4,0) model.  These models have the same one-loop beta functions, 
but differ at two-loop.  Also shown are the theoretical error estimates as discussed in the text.} 
\label{tab:5200}
\end{table}
  
The value of $m_{susy}$ for this solution can be compared to recent bounds on gluinos from the LHC, which now exceed $2$~TeV (for example, see 
Ref.~\cite{ATLAS:2017cjl}).   These bounds generally make assumptions  about the supersymmetric particle spectrum (for example, light neutralinos) 
and one can always play the game of making model-specific adjustments to evade the assumptions of any given experimental exclusion limit.  We
will not pursue that approach here. We instead consider the possibility that $m_V$ and $m_{susy}$ are not identical, so that $m_{susy}$ can be raised unambiguously 
above the LHC reach.  In this case, however, we obtain lower values of $m_V$, which in this model would place an entire vector-like generation below $1$~TeV.   As a point of comparison, current LHC bounds on a charge-$2/3$ vector-like quark that decays 100\% of the time to $bW$ is $1.295$~TeV at the 95\% CL~\cite{CMS:2017wdp}.  
The same comment regarding the limitations of experimental exclusion limits applies here as well;  we will be content simply to point out that the one-generation 
model will become less plausible as time goes on given the increasing reach of LHC searches for superparticles and vector-like quarks.\footnote{Unless, of course, some
of these particles are discovered.}

This result motivates the topic of the next section, extensions of this minimal sector that include sets of new particles that fill complete SU(5) multiplets.  We find that these  
lead to larger values of $m_V$.  In studies of perturbative gauge coupling unification, it is well known that adding additional matter in complete SU(5) multiplets 
preserves successful unification.  In the present framework, we find viable solutions for $m_V$ are also obtained when complete SU(5) multiplets are added.
To study the effect on $m_V$ and $\Lambda$, we consider adding the smallest SU(5) representations, with dimensions five and ten, allowing for 
multiple copies.  We label models by four numbers $(n_g, n_h, n_5, n_{10})$ which represent the number of chiral generations, 
complex Higgs doublets, $\bf{5+\ol 5}$ pairs and $\bf{10+\ol{10}}$ pairs.\footnote{It is interesting to note that in level-one string theories with Wilson line symmetry 
breaking, extra vector-like matter will naturally appear in $\bf{5+\ol 5}$  and $\bf{10+\ol{10}}$ pairs, since these are representations found in the 
${\bf 27+ \ol{27}}$ of $E_6$~\cite{Ghilencea:1997yr}.} In this notation, the one-vector-like-generation scenario that we have discussed in this section
will be called the  $(5,2,0,0)$ model henceforth.   We note that 
a model with four $\bf{5+\ol 5}$ pairs added to the MSSM, the $(3,2,4,0)$ model, has the same one-loop beta functions as the $(5,2,0,0)$ model, 
and could be considered an equally minimal alternative.  Results for the $(3,2,4,0)$ model are also shown in Table~\ref{tab:5200}, and are useful for illustrating the effect 
of different two-loop beta functions. The preferred range of $m_V$ in the $(3,2,4,0)$ model is slightly below that of the $(5,2,0,0)$ model, 
again pointing to the need for alternative choices for the new matter sector to avoid potential phenomenological difficulties.

\section{Next-to-minimal possibilities} \label{sec:next}

In this section, we consider vector-like matter sectors that are consistent with values of $m_{susy}$ and $m_V$ that are no smaller than $2$~TeV.   We 
look at next-to-minimal scenarios, {\em i.e.} ones with a small number of additional $\bd 5+\ol{\bd 5}$ and 
$\bd {10}+\ol{\bd {10}}$ pairs, for the reasons discussed at the end of the previous section.  We have particular interest in solutions that may be plausible 
for exploration at a $100$~TeV hadron collider.  To proceed, we use the results for the one- and two-loop beta functions, derived from the general formulae 
in Refs.~\cite{Yamada:1993ga} and \cite{Machacek:1983tz}.  In the supersymmetric case, we find
\begin{equation}
b_i = \begin{pmatrix}2\\ 2\\ 2\end{pmatrix}n_g+ \begin{pmatrix} \tfrac{3}{10} \\ \tfrac{1}{2} \\ 0\end{pmatrix}n_h + \begin{pmatrix}1\\1\\1 \end{pmatrix}n_5 + \begin{pmatrix}3 \\ 3 \\ 3 \end{pmatrix}n_{10} +\begin{pmatrix}0\\-6\\-9 \end{pmatrix},
\label{eq:olsusy}
\end{equation}
\begin{eqnarray}
 & b_{ij} =\begin{pmatrix} \tfrac{38}{15} & \tfrac{6}{5} & \tfrac{88}{15} \\ \tfrac{2}{5} & 14 & 8 \\ \tfrac{11}{15} & 3 & \tfrac{68}{3} \end{pmatrix} n_g 
+\begin{pmatrix} \tfrac{9}{50} & \tfrac{9}{10} & 0 \\ \tfrac{3}{10} & \tfrac{7}{2} & 0 \\ 0 & 0 & 0 \end{pmatrix} n_h & \nonumber \\
&+\begin{pmatrix} \tfrac{21}{45} & \tfrac{9}{5} & \tfrac{32}{15} \\ \tfrac{3}{5} & 7 & 0 \\ \tfrac{4}{15} & 0 & \tfrac{34}{3} \end{pmatrix} n_5
+\begin{pmatrix} \tfrac{23}{5} & \tfrac{3}{5} & \tfrac{48}{5} \\ \tfrac{1}{5} & 21 & 16 \\ \tfrac{6}{5} & 6 & 34\end{pmatrix} n_{10}
+ \begin{pmatrix} 0 & 0 & 0 \\ 0 & -24 & 0 \\ 0 & 0 & -54\end{pmatrix},&
\label{eq:tlsusy} 
\end{eqnarray}
while in the nonsupersymmetric case,
\begin{equation}
 b_i^{NS}  = \begin{pmatrix} \tfrac{4}{3}  \\  \tfrac{4}{3} \\ \tfrac{4}{3}\end{pmatrix}n_g+
 \begin{pmatrix} \tfrac{1}{10}   \\ \tfrac{1}{6} \\ 0\end{pmatrix}n_h 
 + \begin{pmatrix} \tfrac{2}{3} \\ \tfrac{2}{3}    \\ \tfrac{2}{3} \end{pmatrix}n_5
 + \begin{pmatrix}2  \\ 2  \\ 2 \end{pmatrix}n_{10} 
 +\begin{pmatrix}0 \\-\tfrac{22}{3}\\-11 \end{pmatrix},
 \label{eq:olnonsusy}
\end{equation}
\begin{eqnarray}
& b_{ij}^{NS}  = \begin{pmatrix} \tfrac{19}{15} & \tfrac{3}{5} & \tfrac{44}{15} \\ \tfrac{1}{5} & \tfrac{49}{3} & 4 \\ \tfrac{11}{30} & \tfrac{3}{2} & \tfrac{76}{3} \end{pmatrix} n_g 
+\begin{pmatrix} \tfrac{9}{50} & \tfrac{9}{10} & 0 \\ \tfrac{3}{10} & \tfrac{13}{16} & 0 \\ 0 & 0 & 0 \end{pmatrix} n_h
+\begin{pmatrix} \tfrac{7}{30} & \tfrac{9}{10} & \tfrac{16}{15} \\ \tfrac{3}{10} & \tfrac{49}{6} & 0 \\ \tfrac{2}{15} & 0 & \tfrac{38}{3} \end{pmatrix} n_5 & \nonumber \\
&+\begin{pmatrix} \tfrac{23}{10} & \tfrac{3}{10} & \tfrac{24}{5} \\ \tfrac{1}{10} & \tfrac{49}{2} & 8 \\ \tfrac{3}{5} & 3 & 38\end{pmatrix} n_{10}
+ \begin{pmatrix} 0 & 0 & 0 \\ 0 & -\tfrac{136}{3} & 0 \\ 0 & 0 & -102\end{pmatrix}.&
\label{eq:tlnonsusy}
\end{eqnarray}
As indicated earlier, $n_g, \, n_h, \, n_5 \text{ and } n_{10}$ represent the number of chiral generations, Higgs doublets, $\bf{5+\ol 5}$ and $\bf{10+\ol{10}}$ pairs, respectively.   One can check that these formulae reduce to the expected results for the MSSM, where $n_g = 3, \, n_h = 2, \, n_5 = n_{10}=0$ in Eqs.~(\ref{eq:olsusy}) and (\ref{eq:tlsusy}), and for the standard model, where $n_g = 3, \, n_h=1, \, n_5=n_{10}=0$ in Eqs.~(\ref{eq:olnonsusy}) and~(\ref{eq:tlnonsusy}).

\begin{table}[h!]
\centering
\begin{tabular}{|c|c|c|c|c|c|}
\hline\hline
Model & $m_{\text{SUSY}}$ (TeV) & $m_V$ range (TeV) & $\Lambda$ range (GeV) & $\alpha_3^{-1}(m_Z)\; \%$ error & $\sin^2\theta_W(m_Z)\; \%$ error
\\ \hline 
$(5,2,1,0)$ & 2 & $95-260$ & $4.9 - 8.2 \x 10^{16}$ & $+4.2\%, \; -2.8\%$ & $+1.5\%, \; -1.4\%$
\\ & $m_V$ & $13-28$ & $3.2 - 5.9 \x 10^{16}$ & $+4.0\%, \; -2.7\%$ & $+1.5\%, \; -1.4\%$
\\ \hline
$(3,2,5,0)$ & 2 & $65-217$ & $4.9 - 9.2 \x 10^{16}$ & $+3.4\%, \; -2.2\%$ & $+1.4\%, \; -1.2\%$
\\ & 10 & $17-32$ & $4.1- 5.8 \x 10^{16}$ & $+3.3\%, \; -2.2\%$ & $+1.4\%, \; -1.2\%$
\\ & $m_V$ & $13-17$ & $4.0 -4.9 \x 10^{16}$ & $+3.3\%, \; -2.2\%$ & $+1.4\%, \; -1.2\%$
\\ \hline
$(3,2,6,0)$ & 2 & $3.8- 13 \x 10^3$ & $4.3 - 8.6 \x 10^{16}$ &  $+3.7\%, \; -2.7\%$ & $+1.4\%, \; -1.2\%$
\\ & 10 & $1.2 - 2.6 \x 10^3$ & $3.6 - 5.6 \x 10^{16}$ &  $+3.7\%, \; -2.7\%$ & $+1.4\%, \; -1.2\%$
\\ & 30 & $522-794$ & $3.1-4.0 \x 10^{16}$ &  $+3.6\%, \; -2.7\%$ & $+1.4\%, \; -1.2\%$
\\ \hline
$(3,2,0,2)$ & 2 & $1.6-1.8 \x 10^4$ & $7.1-7.6 \x 10^{16}$ &  $+5.6\%, \; -4.1\%$ & $+1.5\%, \; -1.5\%$
\\ & 10 & $3.0 - 5.3 \x 10^3$ & $4.4 - 6.1 \x 10^{16}$ &  $+5.4\%, \; -3.9\%$ & $+1.5\%, \; -1.5\%$
\\ & 100 & $277 - 961$ & $2.2 - 4.5 \x 10^{16}$ &  $+5.1\%, \; -3.8\%$ & $+1.5\%, \; -1.5\%$
\\ & $m_V$ & $166-370$ & $1.9 - 3.9 \x 10^{16}$ & $+5.0\%, \; -3.7\%$ & $+1.5\%, \; -1.5\%$
\\ \hline\hline
\end{tabular}
\caption{Solutions for $m_V$ and $\Lambda$, for a variety of next-to-minimal heavy matter sectors, for $m_{susy} \leq m_V$.}
\label{tab:newstuff}
\end{table}

Table~\ref{tab:newstuff} displays results analogous to those presented for the minimal scenario in Table~\ref{tab:5200}, for a 
variety of heavy matter sectors, with $m_{susy} \leq m_V$.  The cases considered fall into pairs that have the same one-loop 
beta functions; for example, adding one additional $\bd 5 +\ol{\bd 5}$ pair to the one-vector-like generation scenario gives 
us the $(5,2,1,0)$ model, which has the same $b_i$ as a model with five $\bd 5 +\ol{\bd 5}$ pairs, namely $(3,2,5,0)$. The 
same can be said for the remaining two models, involving six $\bd 5 +\ol{\bd 5}$ and two $\bd {10} +\ol{\bd {10}}$ pairs, 
respectively.  Results are shown for values of $m_{susy}$ ranging from $2$~TeV to $m_V$.  We see that solutions for $m_V$ 
decrease as $m_{susy}$ is increased.  Holding $m_{susy}$ fixed, heavy matter sectors that give larger contributions to the 
one-loop beta functions tend to have larger values of $m_V$.   Larger collections of heavy matter do not provide additional 
solutions with $m_{susy} \leq m_V$ and $m_V < 100$~TeV.

Of the cases shown in Table~\ref{tab:newstuff}, the lowest values of the vector-like matter scale, $m_V \approx 13$~TeV, 
are obtained in the $(5,2,1,0)$ and $(3,2,5,0)$ scenarios, for $m_V = m_{susy}$.  While vector-like quarks with this mass are 
within the kinematic reach of a $100$~TeV hadron collider, their detectability is a separate question.  Assuming
that a $100$~TeV collider has a discovery reach that is greater than that of the LHC by a factor of 5~\cite{Arkani-Hamed:2015vfh}, and 
that the LHC's ultimate sensitivity to vector-like quarks is just below $2$~TeV~\cite{Bhattacharya:2013iea}, one might roughly expect a 
discovery reach for vector-like quarks at a $100$~TeV hadron collider just below $\sim 10$~TeV.  This rough estimate is consistent with the $9$~TeV 
reach projected in Ref.~\cite{Golling:2016gvc} for fermionic top quark partners, which are also color triplet fermions.  These statements are very rough, 
and a detailed collider study would be required to determine whether the $13$~TeV vector-like quarks in the $(5,2,1,0)$ and $(3,2,5,0)$ models would 
have observable consequences at a $100$~TeV machine.

Fortunately, we find that if the supersymmetry-breaking scale is raised above the scale $m_V$, the reduction in $m_V$ continues.
Interestingly, however, we only find the correct predictions for the gauge couplings at the weak scale in the $(3,2,0,2)$ model.   Although a higher $m_{susy}$ indicates that 
supersymmetry is less effective at addressing the hierarchy problem, one could still argue that this case has its merits: {\em (1)} supersymmetry 
still ameliorates the hierarchy problem between $m_{susy}$ and $\Lambda$, which are the scales with the widest separation in the models
that we consider, and {\em (2)} supersymmetry may be expected if string theory is the UV completion, whether or not supersymmetry has anything to do 
with solving the hierarchy problem.  From a purely phenomenological perspective, taking $m_{susy} > m_V$ brings the $(3,2,0,2)$ heavy matter 
sector down into the range where it might be directly probed.  In Table~\ref{tab:mf<msusy}, we present numerical results for that case.   As the supersymmetry 
breaking scale increases from $250$~TeV to $1500$~TeV,  the minimum allowed values of $m_V$ decrease from $71$~TeV to $3$~TeV.  It seems more
likely in this case that the vector-like matter could be within the discovery reach of a $100$~TeV hadron collider, while all the superpartners remain undetectable.  
It is interesting to note that it is easiest in the $(3,2,0,2)$ model to incorporate an additional gauge group that acts on the heavy matter sector, a topic we turn 
to in the next section.

\begin{table}[h!]
\centering
\begin{tabular}{|c|c|c|c|c|c|}
\hline\hline
Model & $m_{\text{SUSY}}$ (TeV) & $m_V$ range (TeV) &  $\Lambda$ range (GeV) & $\alpha_3^{-1}(m_Z)\; \%$ error & $\sin^2\theta_W(m_Z)\; \%$ error
\\ \hline
$(3,2,0,2)$ & 250 & $71-250$ & $1.7 -  2.8 \x 10^{16}$ & $+5.0\%, \; -3.7\%$ & $+1.5\%, \; -1.5\%$
\\ & 500 & $22-216$ & $1.5 -  3.6 \x 10^{16}$ & $+4.9\%, \; -3.6\%$ & $+1.5\%, \; -1.5\%$
\\ & 1000 & $7-64$ & $1.3 -  3.1 \x 10^{16}$ & $+4.8\%, \; -3.5\%$ & $+1.5\%, \; -1.5\%$
\\ & 1500 & $3-31$ & $1.2 -  2.8 \x 10^{16}$ & $+4.7\%, \; -3.5\%$ & $+1.5\%, \; -1.5\%$
\\ \hline\hline
\end{tabular}
\caption{Solutions for $m_V$ and $\Lambda$ for $m_{susy} > m_V$.  Of the models in Table~\ref{tab:newstuff}, only
the $(3,2,0,2)$ case provides viable solutions.}
\label{tab:mf<msusy}
\end{table}

\section{Model building issues} \label{sec:models}
The results of the previous section indicate that there are values of $m_V$ implied by Eq.~(\ref{eq:uvbc}) that are beyond the 
reach of the LHC, but may be within the reach of future collider experiments, particularly in the case where the supersymmetry
breaking scale exceeds the scale $m_V$.   Aside from the extra matter fields, other physics associated with this sector might 
also be experimentally probed.  In this section, we consider two motivations for including an extra gauge group that only affects 
the heavy fields: {\em (1)} The heavy fields may fall in irreducible representations of the new gauge group, explaining the multiplicity of 
new particles required to achieve the blow up of the couplings at the scale $\Lambda$, and {\em (2)} the new sector may be chiral 
under the new gauge groups, rendering it more analogous in structure to the matter sector of the MSSM.   Although there are a large number 
of ways in which either possibility might arise, we consider one example here, based on the $(3,2,0,2)$ model discussed in the previous section.

Regarding the first motivation, we consider the possibility that the duplication of vector-like $\bd {10} +\ol{\bd {10}}$ pairs in
the $(3,2,0,2)$ model is a result of their embedding into a two-dimensional representation of an additional gauge group, 
which is necessarily non-Abelian.   The simplest possibility for the gauge group structure of the model is
$G_{SM} \times $SU(2)$_X$, where $G_{SM}$ represents the standard model gauge factors.  As before, we indicate the standard model charge assignments implicitly and compactly by displaying the SU(5) multiplets that the heavy matter fields would occupy in a conventional unified theory, even though that is not our assumption.  Hence under SU(5)$\times$SU(2)$_X$, we now assume that the extra matter is given by
\begin{equation}
\psi \sim (10,2) \,\,\,\,\, \mbox{ and } \,\,\,\,\, \overline{\psi} \sim (\overline{10},2) \,\,\,.
\label{eq:newmatter}
\end{equation}
We also introduce two SU(2)$_X$ doublet Higgs fields that will be responsible for spontaneously breaking the new gauge group factor
\begin{equation}
\phi_1 \sim (1,2) \,\,\,\,\, \mbox{ and } \,\,\,\,\, \phi_2 \sim (1,2) \,\,\,.
\label{eq:newhiggs}
\end{equation}
The matter fields in Eq.~(\ref{eq:newmatter}) and the new Higgs fields in Eq.~(\ref{eq:newhiggs}) are separately vector-like, so 
that these fields may be made massive at any desired scale; it also follows that all chiral gauge anomalies are canceled.  Note 
that the multiplicity of SU(2) doublets in Eqs.~(\ref{eq:newmatter}) and (\ref{eq:newhiggs}) is even, which implies that the 
SU(2)$_X$ Witten anomaly is absent.  Given these assignments, the one-loop beta function for the new gauge factor is positive, allowing for 
straightforward implementation of the UV boundary condition in Eq.~(\ref{eq:uvbc}).

One issue that needs to be addressed in a model like this one is the stability of the extra matter fields.  Vector-like
$\bd 5 +\ol{\bd 5}$ and $\bd {10} +\ol{\bd {10}}$ pairs have the appropriate electroweak and color quantum numbers to
participate in mass mixing with standard model matter fields.  The amount of such mixing is arbitrary, and only a small
amount is necessary so that the heavy states are rendered unstable, avoiding any cosmological complications.  Assigning
the matter fields of the heavy sector to multiplets of a new gauge group can have unwanted consequences if these states 
are rendered exactly stable (or extremely long lived).  In the present model, this problem does not arise provided that the new
gauge group is spontaneously broken, since mass mixing is generated via renormalizable couplings involving $\psi$, the 
$\phi_i$, and the standard model fields identified with a ${\bf 10}$.  If embedding in an additional gauge group is used to
account for the multiplicity of states in some of the other models that we have considered, the model must also provide
for the decay of the heavy states;  the $(3,2,0,2)$ models seem to naturally avoid this problem with smallest field 
content and the potentially simplest symmetry-breaking sector, which is one reason why we focus on this example here.

Note that the numerical results for the $(3,2,0,2)$ model described in Sec.~\ref{sec:next} must be adjusted to take into
account the presence of the SU(2)$_X$ gauge group, whose coupling blows up at the same scale as the other gauge 
couplings and affects their renormalization group running.  However, since the effect is only via two-loop terms, we don't 
expect a dramatic change in our qualitative conclusions.  To support this statement, we consider the case where 
$m_{susy}=m_V$ and take into account the effect of the new gauge group by modifying the supersymmetric RGEs for 
running between the scales $\Lambda$ and $m_V$.  In this case, the supersymmetric beta functions become
\begin{equation}
b_i =  \left(\begin{array}{cccc} \frac{63}{5} & 7 & 3 & 5 \end{array}\right)  \,\,\, ,
\end{equation}
\begin{equation}
b_{ij} = \left(\begin{array}{cccc} \frac{429}{25} & \frac{33}{5} & \frac{184}{5} & 18 \\
		                            \frac{11}{5}    & 67   & 56 &    18                      \\
		                            \frac{23}{5}  &  21    & 82 &    18                            \\
		                                       6       &  18    & 48  &     53        \end{array}\right)  \,\,\,.
\end{equation}
Repeating the analysis of Sec.~\ref{sec:next}, we find only a modest adjustment in the ranges for $m_V$ and
$\Lambda$, as shown in Table~\ref{table:models} below.

\begin{table}[h!]
\centering
\begin{tabular}{|c|c|c|c|c|c|}
\hline\hline
Model & $m_{\text{SUSY}}$ (TeV) & $m_V$ range (TeV) &  $\Lambda$ range (GeV) & $\alpha_3^{-1}(m_Z)\; \%$ error & $\sin^2\theta_W(m_Z)\; \%$ error
\\ \hline
$(3,2,0,2)$ & $m_V$ & $198-497$ & $1.6 - 3.6 \x 10^{16}$ & $+5.6\%, \; -4.1\%$ & $+1.6\%, \; -1.5\%$
\\ \hline\hline
\end{tabular}
\caption{Results for the $(3,2,0,2)$ scenario with $m_V=m_{susy}$ taking into account the effect of the SU(2)$_X$ gauge group.}
\label{table:models}
\end{table}
                         
It is interesting to note that SU(2)$_X$ breaking scale is not tied to the value of $m_V$ in this model, which means it could in principal 
be much lower.  For example, with $\langle \phi \rangle \sim 1$~GeV, the resulting low-energy effective theory
would be that of a non-Abelian dark sector with a one- or two-Higgs doublet symmetry-breaking sector.  Communication between the visible and 
dark sectors would follow from operators generated when the $m_V$-scale physics is integrated out, suggesting that this sector may have other 
interesting consequences besides its effect on gauge coupling running.  Whether phenomenologically interesting models of this type can be 
constructed remains an open question.  

Finally, we note that a different motivation for an extra gauge factor is to render the $m_V$-scale physics chiral, so that the structure of the 
new matter sector is more similar to the rest of the MSSM.  In the previous example, we could simply change the charge assignment of $\overline{\psi}$ to 
\begin{equation}
\overline{\psi}_1 \sim (\overline{10},1) \,\,\,\,\, \mbox{ and } \,\,\,\,\, \overline{\psi}_2 \sim (\overline{10},1) \,\,\, .
\label{eq:chimod}
\end{equation}
Now the mass terms for the extra matter are generated via Yukawa couplings involving $\psi$, $\overline{\psi}$ and the $\phi_i$; 
the vacuum expectation value $\langle \phi \rangle$ is now associated with the scale $m_V$ determined in the RGE analysis.   We make one additional 
modification to the theory, which is to add an additional pair of Higgs fields
\begin{equation}
\phi'_1 \sim (1,2) \,\,\,\,\, \mbox{ and } \,\,\,\,\, \phi'_2 \sim (1,2) \,\,\,.
\label{eq:xhiggs}
\end{equation}
The modification in Eq.~(\ref{eq:chimod}) leads to the vanishing of the one-loop beta function for 
SU(2)$_X$, while Eq.~(\ref{eq:xhiggs}) restores the desired asymptotic non-freedom.  Based on our earlier observations,
it is clear that the numerical values for $m_V$ and $\Lambda$ in this model will be qualitatively similar to those of the other
$(3,2,0,2)$ models that we have considered, and we leave further numerical study for the interested reader.

\section{Conclusions} \label{sec:conc}
In this paper, we have revisited the possibility that the standard model gauge couplings reach a common Landau pole in
the ultraviolet.  This provides a predictive framework for relating the values of the gauge couplings at the weak scale,
without the necessary assumption of conventional grand unification.  To implement this framework, all the gauge couplings
must be asymptotically non-free, which implies that new matter must be included in the theory.  We have numerically
explored the possibility that this new matter appears at two scales, the scale of supersymmetry breaking, $m_{susy}$, 
and the scale where additional vector-like states appear, $m_V$.  We have revisited a scenario considered
in the past in which the minimal supersymmetric standard model is enlarged by a single vector-like generation and
found that either $m_{susy}$ or $m_V$ falls below potentially relevant LHC lower bounds on colored MSSM superparticles or 
vector-like quarks. Although one cannot rule out the possibility that these states are present and have evaded 
detection for model-specific reasons, we are motivated to consider a safer possibility: we include a relatively 
small additional amount of extra heavy matter, which leads to solutions for $m_V$ that are beyond the reach of the LHC, but potentially
within the reach of a higher-energy hadron collider.  For example, given a heavy sector consisting in total of five $\bd 5 +\ol{\bd 5}$
pairs, we obtain successful gauge coupling predictions for $m_{susy} = m_V \approx 13$~TeV.  For a heavy sector of two
$\bd {10} +\ol{\bd {10}}$ pairs, we can achieve $m_V$ as low as $3$~TeV, if we allow higher values of $m_{susy} \approx 1500$~TeV.

We also considered whether the size of the new matter sector could be related to its embedding into the irreducible 
representation of an additional non-Abelian gauge group.  We presented the simplest model that was consistent with
our numerical solutions, a model with two $\bd {10} +\ol{\bd {10}}$ pairs, in which this duplication is due to their
embedding in the fundamental representation of a new SU(2) gauge group.  In the case where the heavy matter sector
is vector-like under the new SU(2), the new gauge group can be broken at a much lower scale and the effective
theory is that of a spontaneously broken non-Abelian dark sector.  In the case where the heavy matter sector is
chiral under the new SU(2), $m_V$ is associated with the symmetry breaking scale.  In this case, new heavy gauge
bosons would be among the spectrum of particles that might be sought at a future collider with a suitable reach.

\begin{acknowledgments}  
This work was supported by the NSF under Grant PHY-1519644. 
\end{acknowledgments}



\begin{thebibliography}{99}

\bibitem{Langacker:1980js} 
For a review, see P.~Langacker,
``Grand Unified Theories and Proton Decay,''
Phys.\ Rept.\  {\bf 72}, 185 (1981).

\bibitem{Maiani:1977cg} 
L.~Maiani, G.~Parisi and R.~Petronzio,
``Bounds on the Number and Masses of Quarks and Leptons,''
Nucl.\ Phys.\ B {\bf 136}, 115 (1978).
  
\bibitem{Cabibbo:1982hy} 
N.~Cabibbo and G.~R.~Farrar,
 ``An Alternative To Perturbative Grand Unification: How Asymptotically Nonfree Theories Can 
 Successfully Predict Low-energy Gauge Couplings,''
  Phys.\ Lett.\  {\bf 110B}, 107 (1982).

\bibitem{Maiani:1986cp} 
  L.~Maiani and R.~Petronzio,
  ``Low-energy Gauge Couplings and the Mass Gap of $N=1$ Supersymmetry,''
  Phys.\ Lett.\ B {\bf 176}, 120 (1986)
  Erratum: [Phys.\ Lett.\ B {\bf 178}, 457 (1986)].

\bibitem{Moroi:1993zj} 
T.~Moroi, H.~Murayama and T.~Yanagida,
``The Weinberg angle without grand unification,''
Phys.\ Rev.\ D {\bf 48}, R2995 (1993).
 
\bibitem{Andrianov:2013irn} 
A.~A.~Andrianov, D.~Espriu, M.~A.~Kurkov and F.~Lizzi,
``Universal Landau Pole,''
Phys.\ Rev.\ Lett.\  {\bf 111}, no. 1, 011601 (2013).
  
\bibitem{Parisi:1974cf} 
 G.~Parisi,
``On the Value of Fundamental Constants,''
Phys.\ Rev.\ D {\bf 11}, 909 (1975).

\bibitem{Ghilencea:1997yr} 
D.~Ghilencea, M.~Lanzagorta and G.~G.~Ross,
  ``Strong unification,''
  Phys.\ Lett.\ B {\bf 415}, 253 (1997).
  
 \bibitem{Eguchi:1974cg} 
  T.~Eguchi and H.~Sugawara,
  ``Extended Model of Elementary Particles Based on an Analogy with Superconductivity,''
  Phys.\ Rev.\ D {\bf 10}, 4257 (1974).
  
\bibitem{pdg} 
  C.~Patrignani {\it et al.} [Particle Data Group],
  ``Review of Particle Physics,''
  Chin.\ Phys.\ C {\bf 40}, no. 10, 100001 (2016).
  
\bibitem{rge}
 V.~Barger, M.S.~Berger, and P. Ohmann,
 ``Supersymmetric grand unified theories: Two-loop evolution of gauge and Yukawa couplings,''
 Phys.\ Rev.\ D {\bf 47}, 3 (1993).
 
 \bibitem{Dienes:1996du} 
  K.~R.~Dienes,
  ``String theory and the path to unification: A Review of recent developments,''
  Phys.\ Rept.\  {\bf 287}, 447 (1997)
  [hep-th/9602045].

\bibitem{Yamada:1993ga} 
  Y.~Yamada,
  ``Two loop renormalization of gaugino masses in general supersymmetric gauge models,''
  Phys.\ Rev.\ Lett.\  {\bf 72}, 25 (1994)
  [hep-ph/9308304].
  
\bibitem{Machacek:1983tz} 
  M.E.~Machacek and M.~T.~Vaughn,
  ``Two Loop Renormalization Group Equations in a General Quantum Field Theory. 1. Wave Function Renormalization,''
  Nucl.\ Phys.\ B {\bf 222}, 83 (1983).

  
  \bibitem{Antoniadis:1982qw} 
   I.~Antoniadis, C.~Kounnas and R.~Lacaze,
  ``Light Gluinos in Deep Inelastic Scattering,''
  Nucl.\ Phys.\ B {\bf 211}, 216 (1983).
 
  \bibitem{g1renorm} 
  M.~E.~Peskin,
  ``Beyond the standard model,''
  In {\em Carry-le-Rouet 1996, High-energy physics} 49-142
  [hep-ph/9705479].
  
 \bibitem{ATLAS:2017cjl} 
  The ATLAS collaboration [ATLAS Collaboration],
  ``Search for squarks and gluinos in final states with jets and missing transverse momentum using 36 fb$^{-1}$ of $\sqrt{s} =13$ TeV pp collision data with the ATLAS detector,''
  ATLAS-CONF-2017-022.
  
  \bibitem{CMS:2017wdp} 
  CMS Collaboration [CMS Collaboration],
  ``Search for vector-like quark pair production $\mathrm{T\bar{T}}(\mathrm{Y\bar{Y}})\rightarrow\mathrm{bWbW}$ using kinematic reconstruction in lepton+jets final states at $\sqrt{s}$=13 TeV,''
  CMS-PAS-B2G-17-003.
  
  \bibitem{Arkani-Hamed:2015vfh} 
  N.~Arkani-Hamed, T.~Han, M.~Mangano and L.~T.~Wang,
  ``Physics opportunities of a 100 TeV proton?proton collider,''
  Phys.\ Rept.\  {\bf 652}, 1 (2016)
  [arXiv:1511.06495 [hep-ph]].
  
  \bibitem{Bhattacharya:2013iea} 
  S.~Bhattacharya, J.~George, U.~Heintz, A.~Kumar, M.~Narain and J.~Stupak,
  ``Prospects for a Heavy Vector-Like Charge 2/3 Quark T search at the LHC with $\sqrt{s}=14$~TeV and 33 TeV. "A Snowmass 2013 Whitepaper",''
  arXiv:1309.0026 [hep-ex].
  
  \bibitem{Golling:2016gvc} 
  T.~Golling {\it et al.},
  ``Physics at a 100 TeV pp collider: beyond the Standard Model phenomena,''
  Submitted to: Phys.\ Rept.
  [arXiv:1606.00947 [hep-ph]].
  
  

\end{thebibliography}
\end{document}